\documentstyle[12pt,psfig,draft,lscape,array,amssymb]{nature-ejb}

\def\simlt{\mathrel{\hbox{\rlap{\hbox{\lower4pt\hbox{$\sim$}}}\hbox{$<$}}}}
\def\simgt{\mathrel{\hbox{\rlap{\hbox{\lower4pt\hbox{$\sim$}}}\hbox{$>$}}}}

\def\ale{\mathrel{\hbox{\rlap{\hbox{\lower4pt\hbox{$\sim$}}}\hbox{$<$}}}}
\def\age{\mathrel{\hbox{\rlap{\hbox{\lower4pt\hbox{$\sim$}}}\hbox{$>$}}}}

\def\nodata{---}

\def\ra#1#2#3{#1$^{\rm h}$#2$^{\rm m}$#3$^{\rm s}$}
\def\dec#1#2#3{$#1^\circ#2'#3''$}

\def\grb{GRB\,031203}

\def\spose#1{\hbox to 0pt{#1\hss}}
\newcommand\lsim{\mathrel{\spose{\lower 3pt\hbox{$\mathchar"218$}}
     \raise 2.0pt\hbox{$\mathchar"13C$}}}
\newcommand\gsim{\mathrel{\spose{\lower 3pt\hbox{$\mathchar"218$}}
     \raise 2.0pt\hbox{$\mathchar"13E$}}}

\begin{document}

\title{\Large \bf The sub-energetic GRB\,031203 as a cosmic
analogue to GRB\,980425}

\author{
   A.~M.~Soderberg\affiliation[1]
     {Caltech Optical Observatories 105-24, California Institute of 
     Technology, Pasadena, CA\,91125, USA \vspace{0.1in}},
   S.~R.~Kulkarni\affiliationmark[1],
   E.~Berger\affiliationmark[1],
   D.~W.~Fox\affiliationmark[1],
   M.~Sako\affiliation[2]
     {Stanford Linear Accelerator Center, 2575 Sand Hill Road M/S 29, Menlo Park, CA\,94025, USA \vspace{0.1in}},
   D.~A.~Frail\affiliation[3]
     {National Radio Astronomy Observatory, P.O. Box 0, Socorro, New
     Mexico 87801, USA \vspace{0.1in}},
   A.~Gal-Yam\affiliationmark[1],
   D.~S.~Moon\affiliation[4]
     {Space Radiation Laboratory 220-47, California Institute of 
     Technology, Pasadena, CA\,91125, USA \vspace{0.1in}}, 
   S.~B.~Cenko\affiliationmark[4],
   S.~A.~Yost\affiliationmark[4],
   M.~M.~Phillips\affiliation[5]
     {Carnegie Observatories, 813 Santa Barbara Street, Pasadena, 
     CA\,91101, USA \vspace{0.1in}},
   S.~E.~Persson\affiliationmark[5],
   W.~L.~Freedman\affiliationmark[5],
   P.~Wyatt\affiliationmark[5],
   R.~Jayawardhana\affiliation[6]
     {Department of Astronomy, University of Michigan, 830 Dennison Bldg, Ann Arbor MI\,48109 USA \vspace{0.1in}} \&\
   D.~Paulson\affiliationmark[6]
}

\date{\today}{}
\headertitle{Sub-energetic GRB\,031203}
\mainauthor{Soderberg et al.}

\summary{Over the six years since the discovery\cite{paa+00} of the
$\gamma$-ray burst GRB\,980425, associated\cite{gvv+98} with the
nearby (distance, $\sim 40$ Mpc) supernova 1998bw, astronomers have
fiercely debated the nature of this event.  Relative to bursts located
at cosmological distances, (redshift, $z\sim 1$), GRB\,980425 was
under-luminous in $\gamma$-rays by three orders of magnitude.  Radio
calorimetry\cite{kfw+98,lc99} showed the explosion was sub-energetic
by a factor of 10.  Here, we report observations of the radio and
X-ray afterglow of the recent $z=0.105$
GRB\,031203\cite{gmb+03,pbc+04,sls04} and demonstrate that it too is
sub-energetic.  Our result, when taken together with the low
$\gamma$-ray luminosity\cite{sls04}, suggest that GRB\,031203 is the
first cosmic analogue to GRB\,980425. We find no evidence that this
event was a highly collimated explosion viewed off-axis.  Like
GRB\,980425, GRB\,031203 appears to be an intrinsically sub-energetic
$\gamma-$burst. Such sub-energetic events have faint afterglows.
Intensive follow-up of faint bursts with smooth $\gamma-$ray light
curves\cite{bkh+98,nor02} (common to both GRBs 031203 and 980425)
may enable us to reveal their expected large population.}

\maketitle


On 3 December 2003 at 22:01:28 UT, the {\it INTEGRAL} satellite
detected\cite{gmb+03,sls04} a seemingly typical long-duration ($\Delta
t\approx 20$ sec) $\gamma$-ray burst.  Within 6 hours, the {\it Newton
X-ray Multiple Mirror (XMM)} Observatory 
detected\cite{scg+03,whl+04} an X-ray source with flux (2--10 keV band),
$F_{X}=3.95\pm 0.09 \times 10^{-13}~\rm erg~cm^{-2}~s^{-1}$, fading
gradually $\propto t^\alpha$ with $\alpha=-0.4$.  Using the Very Large
Array (VLA), we discovered a radio source at
$\alpha$(J2000)=\ra{08}{02}{30.18} and
$\delta$(J2000)=\dec{-39}{51}{03.51} ($\pm 0.1$ arcsec in each axis),
well within the 6-arcsecond radius error circle of the {\it XMM}
source.  A subsequent {\it XMM} observation\cite{rsg+03} confirmed the
gradual decay of the X-ray source.  From our analysis of the {\it XMM}
data we find the flux $\propto t^{-0.4}$ between the two epochs and the
spectral flux density, $F_{\nu, X} \propto \nu^{\beta}$, is fit by
$\beta=-0.81\pm 0.05$ with an absorbing column density, $N_{\rm H} =
6.2\times 10^{21}~\rm cm^{-2}$. Taken together, the transient X-ray and
radio emission are suggestive of afterglow emission.

In addition to monitoring the afterglow in various radio bands
(Table~\ref{tab:vla} and discussion below) we obtained an observation
of the source with the {\it Advanced CCD Imaging Spectrometer} (ACIS)
instrument aboard the {\it Chandra} X-ray Observatory ({\it CXO}).  The
{\it CXO} observations began on 22 January 2004 at 21:35 UT and lasted
about 22 ksec. We detected a faint source, count rate in the 2--10 keV
band of $5.6\times 10^{-4}\,$s$^{-1}$, at
$\alpha$(J2000)=\ra{08}{02}{30.159} and
$\delta$(J2000)=\dec{-39}{51}{03.51} ($\pm 0.18$ arcsec in each
axis), precisely coincident with the VLA source. Using the
{\it XMM} model parameters stated above we obtain $F_X= 6.4\times
10^{-15}~\rm erg~cm^{-2}~s^{-1}$, implying a faster decline
($\alpha=-1\pm 0.1$) between the second {\it XMM} and {\it Chandra}
observations.

The primary interest in this burst is that the radio and X-ray
afterglow coincides at the sub-arcsecond level \cite{gal04} with a
nearby ($z=0.1055$) galaxy \cite{pbc+04}, making it the nearest GRB
with the exception of the peculiar GRB\,980425\cite{gvv+98}.  At this
redshift, the isotropic $\gamma-$ray energy release is $10^2$ times
smaller\cite{sls04} than that of the nearest classical event
GRB\,030329 ($z=0.169$)\cite{vsb+04} and yet a factor of $10^2$
larger\cite{paa+00,gvv+98} than that of GRB\,980425.

The afterglow properties of GRB\,031203 also appear to be intermediate
between classical cosmological GRBs and GRB\,980425: the isotropic
X-ray luminosity of GRB\,031203 at $t\approx 10$ hours is
$L_{X}=9\times 10^{42}~\rm erg~cm^{-2}~s^{-1}$, nearly $10^3$ times
fainter than that observed\cite{bkf03} for classical GRBs but a factor
of $10^2$ brighter\cite{paa+00} than that of GRB\,980425.  In the
centimetre band, the peak luminosity is $L_{\nu, 8.5~\rm GHz}\approx
10^{29}~\rm erg~s^{-1}~Hz^{-1}$, fainter\cite{fkb+03} by a factor of
$10^{2}$ than that of most radio afterglows but comparable\cite{kfw+98}
to that of GRB\,980425.  Since $L_X$
and peak radio luminosity of an afterglow can be
used\cite{pk02,bkf03} as rough proxies for the afterglow energy, the
data suggest that GRBs 031203 and 980425 are sub-energetic in
comparison with classical GRBs.

\begin{figure}
\centerline{\psfig{file=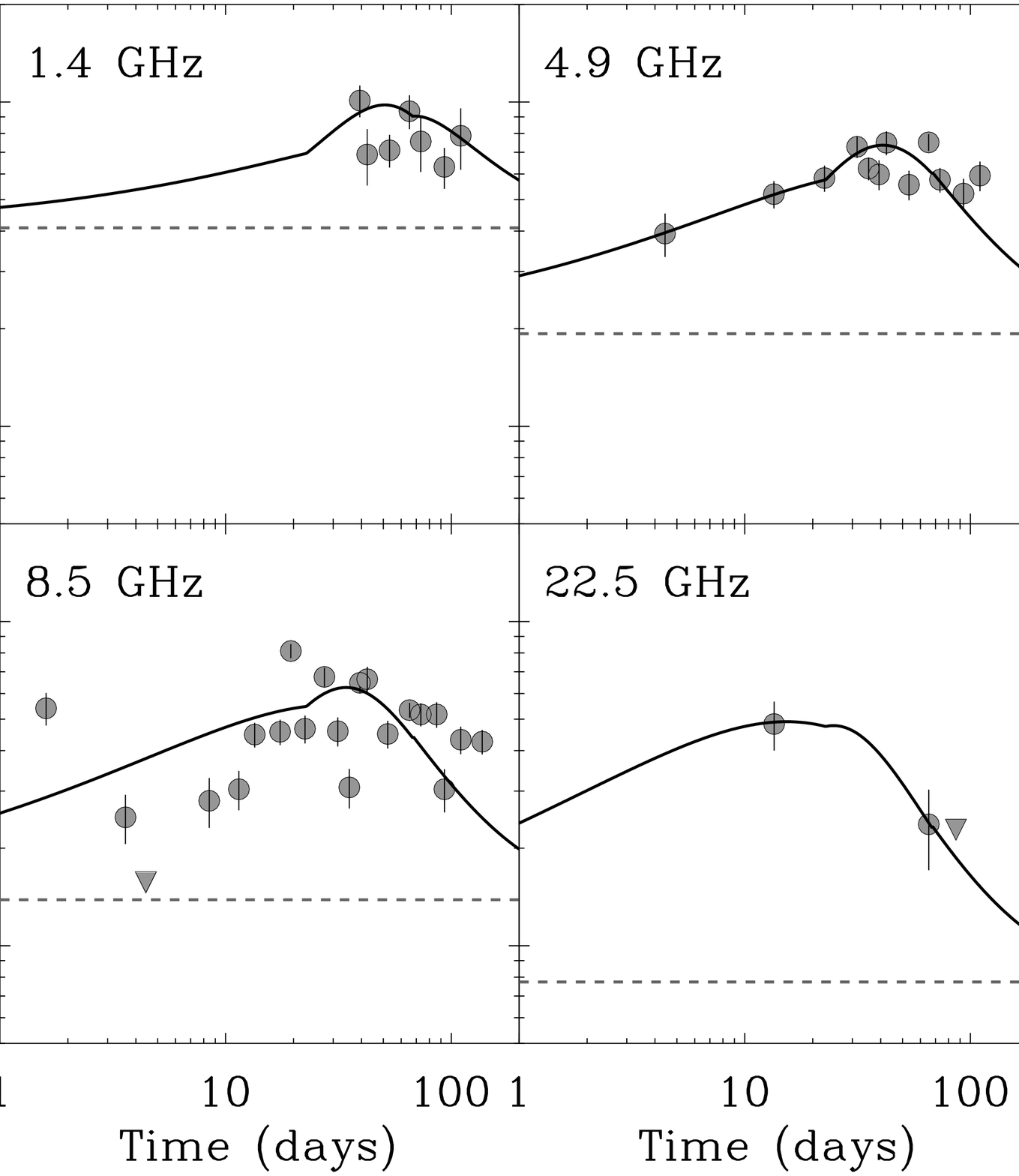,width=5.in,angle=0}}
\bigskip
\bigskip
\caption[]{}
\label{fig:LightCurves}
\end{figure}

As a next step, we applied the simplest afterglow
model\cite{spn98,gs02} (a spherical relativistic blastwave shocking a
constant density circumburst medium and accelerating relativistic
electrons; the afterglow emission arises from synchrotron emission of
shocked electrons) to the afterglow data and obtain a satisfactory fit
(Figure~\ref{fig:LightCurves}).  On the timescales best probed by the
radio data -- days to months -- we see no evidence for a collimated
(jet) geometry commonly seen\cite{fks+01} in the afterglows of
cosmological GRBs.

From our modeling we confirm that the blast wave is sub-energetic, finding 
an inferred afterglow energy of $E_{\rm AG}\approx 1.7 \times 10^{49}$ erg.  
The circumburst particle density $n\approx 0.6\,{\rm cm^{-3}}$, is not
atypical of that inferred\cite{pk02} for other GRBs.
The blastwave is expected to become\cite{wax03} non-relativistic on a
timescale, $t_{\rm NR}\approx 34(E_{\rm AG,50}/n_0)^{1/3}\,$d, where
we adopt the notation $q\equiv 10^x q_x$.  The observational
signatures\cite{fwk00} of this transition, a steeper decay of the
spectral peak frequency ($\nu_m \propto t^{-1.5} \rightarrow t^{-3}$)
and an increase in the spectral peak flux ($F_{\nu_m} \propto t^0
\rightarrow t^{3/5}$) are consistent with the data
(Figure~\ref{fig:LightCurves}).

\begin{figure}
\centerline{\psfig{file=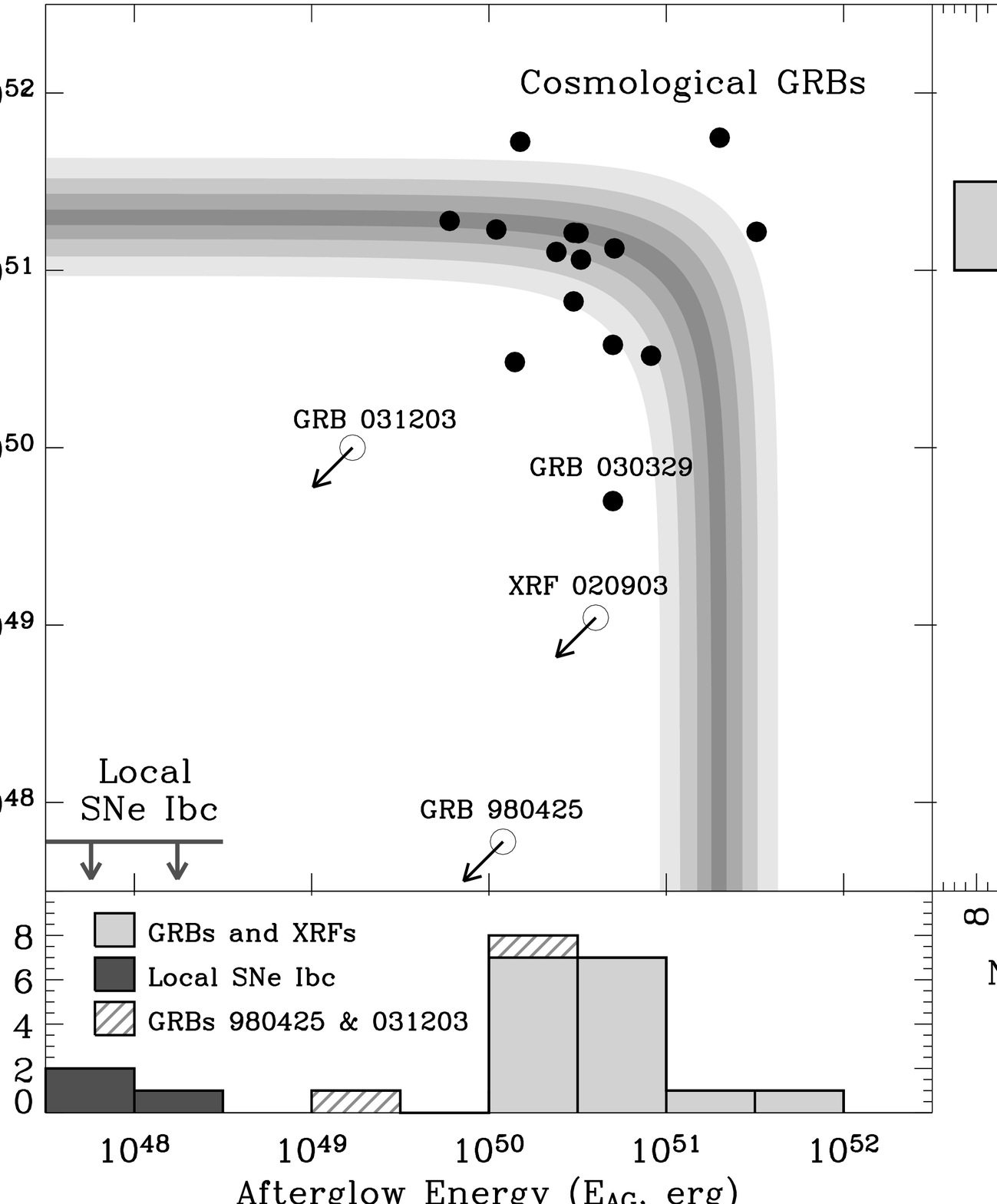,width=5in,angle=0}}
\bigskip
\bigskip
\caption[]{}
\label{fig:EnergyHistogram}
\end{figure}

Here we use $E_{\rm AG}$ to denote the kinetic energy remaining in the
blast wave after the prompt $\gamma-$ray energy release.  In turn,
the $\gamma$-ray emission arises from ultra-relativistic (bulk Lorentz
factor, $\Gamma\age 100$) ejecta within the blastwave.  Thus, a more
complete picture of the explosion energy is visualized through a 
two-dimensional plot of $E_{\rm prompt}$, the beaming-corrected prompt 
energy release versus $E_{\rm AG}$ (Figure~\ref{fig:EnergyHistogram}).

The two nearest events, GRBs 031203 and 980425, are clearly sub-energetic
outliers in Figure~\ref{fig:EnergyHistogram}.  Furthermore, we draw
the reader's attention to several additional similarities: GRBs 031203 and
980425 (1) show no evidence for jets\cite{kfw+98}, (2) possess simple
$\gamma$-ray light curves\cite{paa+00,sls04}; and with respect to
cosmological (``classical'') bursts the two events (3)
violate\cite{sls04} the $E_{\rm prompt}-E_{\rm peak}$
relation\cite{aft+02} and (4) are outliers in the luminosity -
spectral lag relation\cite{nor02}.  This discussion motivates the question:
{\it How are these two events related to cosmological GRBs?}

It has been suggested (e.g.~ref~\pcite{nmn+01}) that all GRB
explosions have the same energetics and explosion geometry.  In this framework,
sub-energetic bursts are simply events viewed away from the jet
axis. Such bursts should have a soft $E_{\rm peak}$ and also exhibit a
rise in the inferred $E_{\rm AG}$ as shocked ejecta eventually come
into our line of sight.  For GRB\,031203, $E_{\rm peak}>190\,$keV
(ref.~\pcite{sls04}), comparable to cosmological GRBs for which we
have observational evidence favoring an on-axis viewing angle.
Moreover, we see no evidence for an increase in $E_{\rm AG}$ during
the timescale of the radio observations ($\sim 150$ days).  Similarly,
there is no evidence that $E_{\rm AG}$ is increasing for GRB\,980425
despite dedicated radio monitoring\cite{sfw04} of the source since 1998.  
With no
indication of being off-axis explosions, we presently conclude that
GRBs 031203 and 980425 are intrinsically sub-energetic events.

Astronomers have had to wait six years to discover a sub-energetic
event similar to GRB\,980425, despite a large population (as implied
by their proximity).  The bulk of the population has escaped our
attention due to their faint $\gamma$-ray and afterglow emission which
challenge our current detection limits.  The {\em Swift} satellite
mission, with its higher $\gamma$-ray sensitivity (compared to current
missions) and improved localization capability (enabling rapid
identification of afterglow counterparts) is expected to revolutionize
our understanding of cosmic explosions.

\noindent
Correspondence should be addressed to A. M. Soderberg
(e-mail:ams@astro.caltech.edu).

\begin{acknowledge}
GRB research at Caltech is supported in part by funds from NSF and
NASA.  We are, as always, indebted to Scott Barthelmy and the GCN.
The VLA is operated by the National Radio Astronomy Observatory, a
facility of the National Science Foundation operated under cooperative
agreement by Associated Universities, Inc..  AMS is supported by an NSF
Graduate Research Fellowship. AG acknowledges support 
by NASA through a Hubble Fellowship grant.
\end{acknowledge}

\clearpage
\begin{table}
\begin{center}
\setlength{\extrarowheight}{-0.075in}
\begin{tabular}{>{\scriptsize}l >{\scriptsize}c >{\scriptsize}c >{\scriptsize}c >{\scriptsize}c >{\scriptsize}c}
\hline
\hline
Epoch & $\Delta t$ & $F_{1.43}$ & $F_{4.86}$ & $F_{8.46}$ &
$F_{22.5}$ \\
(UT) & (days) & (mJy) & (mJy) & (mJy) & (mJy) \\
\hline 
2003 Dec 5.52  & 1.60 & \nodata & \nodata & $0.540\pm 0.062$ & \nodata \\
2003 Dec 7.52  & 3.60 & \nodata & \nodata & $0.249\pm 0.043$ & \nodata \\
2003 Dec 8.35  & 4.43 & \nodata & $0.393\pm 0.060$ &  $0.053\pm 0.052$ & \nodata \\
2003 Dec 12.38 & 8.46 & \nodata & \nodata & $0.280\pm 0.049$ & \nodata \\
2003 Dec 15.37 & 11.45 & \nodata & \nodata & $0.304\pm 0.042$ & \nodata \\
2003 Dec 17.38 & 13.46 & \nodata & $0.520\pm 0.050$ &  $0.448\pm 0.039$ & $0.483\pm  0.083$ \\
2003 Dec 21.35 & 17.43 & \nodata & \nodata & $0.457\pm 0.041$ & \nodata \\
2003 Dec 23.37 & 19.45 & \nodata & \nodata & $0.811\pm 0.040$ & \nodata \\
2003 Dec 26.40 & 22.48 & \nodata & $0.583\pm 0.054$ &  $0.467\pm 0.046$ & \nodata \\
2003 Dec 31.33 & 27.41 & \nodata & \nodata & $0.675\pm 0.045$ & \nodata \\
2004 Jan 4.33  & 31.41 & \nodata & $0.728\pm 0.055$ &  $0.459\pm  0.047$ & \nodata \\
2004 Jan 8.26 & 35.34 & \nodata & $0.624\pm 0.050$ &  $0.308\pm 0.043$ & \nodata \\
2004 Jan 12.29 & 39.37 & $1.011\pm 0.113$ & $0.598\pm 0.063$ &  $0.647\pm 0.045$ & \nodata \\
2004 Jan 15.35 & 42.43 & $0.689\pm 0.136$ & $0.749\pm 0.063$ &  $0.664\pm 0.061$ & \nodata \\
2004 Jan 25.24 & 52.32 & $0.710\pm 0.082$ & \nodata &  $0.450\pm 0.044$ & \nodata \\
2004 Jan 26.34 & 53.42 & \nodata & $0.556\pm 0.058$ & \nodata & \nodata \\
2004 Feb 7.24  & 65.32 & $0.937\pm 0.112$ & $0.751\pm 0.045$ & $0.533\pm 0.028$ & $0.273\pm0.066$ \\
2004 Feb 15.22 & 73.30 & $0.756\pm 0.147$ & $0.576\pm 0.050$ & $0.517\pm 0.042$ & \nodata \\
2004 Feb 28.13 & 86.21 & \nodata & \nodata & $0.517\pm 0.047$ & $0\pm 0.114$ \\
2004 Mar 6.17 & 93.25 & $0.631\pm 0.091$ & $0.522\pm 0.058$ & $0.304\pm 0.046$ & \nodata \\
2004 Mar 23.13 & 110.21 & $0.787\pm 0.169$ & $0.593\pm 0.062$ &  $0.432\pm 0.042$ & \nodata \\	
2004 Apr 19.07 & 137.15 & \nodata & \nodata & $0.426\pm 0.037$ & \nodata \\
\end{tabular}
\end{center}
\caption[]{\small Radio observations made with the Very Large Array 
(VLA).  Observations commenced on 5 December 2003 UT.  
For all observations we used the standard continuum
mode with $2\times 50$ MHz bands.  At 22.5 GHz we used
referenced pointing scans to correct for the systematic $10 - 20$ arcsec
pointing errors of the VLA antennas.  We used the extra-galactic
sources 3C\,147 (J0542+498) and 3C\,286 (J1331+305) for flux
calibration, while the phase was monitored using J0828-375.
The data were reduced and analyzed using the Astronomical 
Image Processing System. 
The flux density and uncertainty were measured from the
resulting maps by fitting a Gaussian model to the afterglow emission.  }
\label{tab:vla}
\end{table}

\clearpage

\noindent
{\bf Figure 1}: {\small Radio lightcurves of the afterglow of \grb{}.
All measurements are summarized in Table~\ref{tab:vla} and include
$1-\sigma$ error bars.  The solid lines are models of synchrotron
(afterglow) emission from spherical ejecta expanding into a uniform
circumburst medium\cite{gs02}.  The models include a contribution from
the host galaxy, which is well-fit by $F_{\rm host}\approx
0.4(\nu/1.4~\rm GHz)^{-0.6}$ mJy (dashed lines) and is consistent with
the star formation rate inferred\cite{pbc+04} from optical
spectroscopy of the host.  In applying the models, the X-ray
observations are considered upper-limits since they are most likely
dominated by (non-synchrotron) emission arising from the associated
supernova SN\,2003lw, as evidenced by the unusually slow flux decay at
early time and the flat spectral index ($F_{\nu ,X}\propto
t^{-0.4}\nu^{-0.8}$ as opposed to $\propto t^{-1}\nu^{-1.3}$ for
GRBs). This was also the case for the X-ray emission\cite{paa+00} of
GRB\,980425/SN\,1998bw ($F_{\nu ,X}\propto t^{-0.2}\nu^{-1}$).  For
our best-fit model, we find $\chi^2_r=8.9$ (38 degrees of freedom),
dominated by interstellar scintillation.  The blastwave transitions to
the non-relativistic regime at $t_{NR}\approx 23$ d.  From the derived
synchrotron parameters (at $t=1$ d): $\nu_a\approx 3.2\times 10^{8}$
Hz, $\nu_m\approx 3.6\times 10^{12}$ Hz and $F_{\nu_a}\approx 0.04$
mJy we find an isotropic afterglow energy, $E_{\rm AG,~iso}\approx
1.7\times10^{49}~\nu_{c,15.5}^{1/4}$ erg, a circumburst density
$n\approx 0.6~\nu_{c,15.5}^{3/4}~\rm cm^{-3}$ and the fractions of
energy in the relativistic electrons (energy distribution $N(\gamma)
\propto \gamma^{-p}$ with $p\approx 2.6$) and magnetic field of
$\epsilon_e\approx 0.4~\nu_{c,15.5}^{1/4}$ and $\epsilon_B\approx
0.2~\nu_{c,15.5}^{-5/4}$, respectively.  Here, $\nu_c=3\times
10^{15}~\nu_{c,15.5}$ is the synchrotron cooling frequency which 
is roughly constrained by the (non-synchrotron) SN\,2003lw X-ray emission.
Extrapolation of the synchrotron model beyond $\nu_c$ underestimates
the observed X-ray flux by a factor of $\lesssim 10$ which
is comparable to the discrepancy for SN\,1998bw (found by
extrapolating the radio model by Li $\&$ Chevalier\cite{lc99}
($p=2.5$, $\epsilon_B=10^{-3}$) beyond $\nu_c$ and comparing with the
X-ray data\cite{paa+00} at $t\sim 12$ days).}

\bigskip

\noindent
{\bf Figure 2}: {\small Two-dimensional energy plot for 
cosmic explosions.  The energy in the prompt emission, $E_{\rm prompt}$, 
and in the afterglow, $E_{\rm AG}$, have been
corrected\cite{fks+01,bfk03,bkp+03} for beaming based on the jet-break
time observed for each burst, except in the cases of
GRB\,980425\cite{lc99,kfw+98}, XRF\,020903\cite{skb+03} and
GRB\,031203 for which there is no evidence for a collimated outflow.
For these three cases we plot the isotropic values of $E_{\rm prompt}$
and $E_{\rm AG}$ and use an arrow to indicate they represent upper
limits on both axes.  The arcs mark lines of constant
$E_{\rm prompt}+E_{\rm AG}$ as a guide to the reader.
Most cosmological GRBs tend to cluster\cite{bkp+03} around $E_{\rm 
prompt}+E_{\rm AG}\approx 2\times 10^{51}$ erg while GRBs 031203 and 
980425, the nearest two bursts in the sample, are clearly sub-energetic.
With the exception of SN\,1998bw, associated with GRB\,980425, there
are no local Ibc supernovae with detected $\gamma$-ray emission, 
however the kinetic energy in the ejecta (excluding the photospheric
energy yield), is generally found\cite{bkfs03} to be $E_{\rm AG} \lesssim 
3\times 10^{48}$ erg (bottom left corner). Histograms of $E_{\rm AG}$ and 
$E_{\rm prompt}$ are shown in the bottom and side panels, respectively, 
for cosmological GRBs and local Ibc SNe.  The striped energy bins show the
locations of GRBs 980425 and 031203.}

\clearpage

\clearpage

\end{document}